\pdfoutput=1
\documentclass[preprint,preprintnumbers,amsmath,amssymb,floatfix,endfloats*]{revtex4}

\usepackage{graphicx}
\usepackage{dcolumn}
\usepackage{bm}
\usepackage{amsfonts}

\DeclareMathAlphabet{\mathsfsl}{OT1}{cmr}{bx}{it}
\begin{document}
\title{The effect of thermal history on the atomic structure and mechanical properties of amorphous alloys}
\author{Nikolai V. Priezjev$^{1,2}$}
\affiliation{$^{1}$Department of Mechanical and Materials
Engineering, Wright State University, Dayton, OH 45435}
\affiliation{$^{2}$National Research University Higher School of
Economics, Moscow 101000, Russia}
\date{\today}
\begin{abstract}

The influence of thermal processing on the potential energy, atomic
structure, and mechanical properties of metallic glasses is examined
using molecular dynamics simulations. We study the three-dimensional
binary mixture, which was first relaxed near the glass transition
temperature, and then rapidly cooled deep into the glass phase.  It
was found that glasses prepared at higher annealing temperatures are
relocated to higher energy states and their average glass structure
remains more disordered, as reflected in the shape of the pair
correlation function. The results of mechanical testing demonstrate
that both the shear modulus and yielding peak increase significantly
when the annealing temperature approaches $T_g$ from above.
Moreover, the shear modulus becomes a strong function of strain rate
only for samples equilibrated at sufficiently high temperatures.
Based on the spatial distribution of nonaffine displacements, we
show that the deformation mode changes from brittle to ductile upon
increasing annealing temperature.  These results can be useful for
the design and optimization of the fabrication processes of bulk
glassy alloys with improved plasticity.

\vskip 0.5in


\end{abstract}

\maketitle

\section{Introduction}

Recent progress in the thermal treatment and deformation processing
of amorphous alloys has resulted in a wider range of accessible
energy states and optimized physical and mechanical
properties~\cite{Qiao19}.  It is well known that metallic glasses
are much stronger than crystalline materials due to the absence of
topological defects, or dislocations, but their deformation often
proceeds via the formation of the so-called shear bands.  Generally,
aged samples become more brittle, while rejuvenated, or higher
energy glasses, are more ductile.   Common methods to rejuvenate
metallic glasses and improve their plasticity include cold rolling,
high pressure torsion, ion irradiation, and shot
peening~\cite{Greer16}.  Interestingly, an application of
elastostatic loading over extended time intervals can significantly
rejuvenate metallic glasses due to plastic deformation of relatively
soft
domains~\cite{Park08,Tong13,GreerSun16,Zhang17,PanGreer18,Samavatian19,PriezELAST19}.
A novel method to control rejuvenation consists of heating up
metallic glasses slightly above the glass transition temperature,
followed by rapid quenching back to the glass
phase~\cite{Ogata15,Maass18,Yang18,Priez19one}. More recently, it
was found that cryogenic thermal cycling might lead to higher energy
states via local plastic events that are triggered because of the
heterogeneity in the local thermal
expansion~\cite{Ketov15,Barrat18,Priez19tcyc,Priez19T2000,Guo19}. In
addition, it was demonstrated that the atomic structure and internal
energy of amorphous materials can be tuned by cyclic
deformation~\cite{Priezjev13,Sastry13,Reichhardt13,Priezjev14,IdoNature15,
Priezjev16,Priezjev16a,Sastry17,Priezjev17,Priezjev18,Priezjev18a,Lo10,NVP18strload}.
Finally, recent experimental studies showed that the
ductile-to-brittle transition at room temperature is controlled by
the so-called fictive temperature, which characterizes the average
glass structure during a rapid quench below the glass transition
temperature~\cite{Schroers13,Schroers18}.   However, despite
extensive efforts, the effect of thermo-mechanical processing on
microstructure and mechanical properties of amorphous alloys remains
not fully understood.

\vskip 0.05in

In this paper, we investigate the dependence of the potential
energy, atomic structure and mechanical properties of binary glasses
on the preparation protocol that consists of equilibration at an
annealing temperature near the glass transition point and a
subsequent quench to the test temperature.  It will be shown that
glasses prepared at lower annealing temperatures settle down at
deeper potential energy minima and their short-range structure
becomes more ordered. As a result, the shear modulus and stress
overshoot depend strongly on the annealing temperature.  We find
that the rate dependence of the shear modulus becomes significant
when the annealing temperature is greater than the glass transition
temperature. The analysis of nonaffine displacements indicates that
upon increasing annealing temperature, deformation changes from
being localized within a shear band to more homogeneous.

\vskip 0.05in

This paper is organized as follows. The description of the molecular
dynamics simulation model as well as the preparation and deformation
protocols is provided in the next section.  The dependence of the
potential energy, radial distribution function, mechanical
properties, and distribution of nonaffine displacements on the
annealing temperature is presented in section\,\ref{sec:Results}. A
brief summary is given in the last section.

\section{Molecular dynamics simulation model}
\label{sec:MD_Model}

In this work, the structure and mechanical properties of an
amorphous alloy are investigated by means of molecular dynamics
simulations. We consider a well-studied binary (80:20) mixture
originally introduced by Kob and Andersen (KA) about two decades
ago~\cite{KobAnd95}. The interaction parameters of the KA mixture
are similar to the parametrization used by Weber and Stillinger to
study the amorphous metal alloy
$\text{Ni}_{80}\text{P}_{20}$~\cite{Weber85}. More specifically, the
atoms of types $\alpha,\beta=A,B$ in the KA model interact via the
Lennard-Jones (LJ) potential:
\begin{equation}
V_{\alpha\beta}(r)=4\,\varepsilon_{\alpha\beta}\,\Big[\Big(\frac{\sigma_{\alpha\beta}}{r}\Big)^{12}\!-
\Big(\frac{\sigma_{\alpha\beta}}{r}\Big)^{6}\,\Big],
\label{Eq:LJ_KA}
\end{equation}
with the following LJ parameters: $\varepsilon_{AA}=1.0$,
$\varepsilon_{AB}=1.5$, $\varepsilon_{BB}=0.5$, $\sigma_{AA}=1.0$,
$\sigma_{AB}=0.8$, $\sigma_{BB}=0.88$, and
$m_{A}=m_{B}$~\cite{KobAnd95}.   Note that this parametrization
represents a system with strongly non-additive interaction between
different types of atoms, which suppresses crystallization near the
glass transition temperature~\cite{KobAnd95}. The total number of
atoms is $N=60\,000$. To alleviate the computational load, the LJ
potential was truncated at the cutoff radius
$r_{c,\,\alpha\beta}=2.5\,\sigma_{\alpha\beta}$.   In what follows,
all physical quantities are expressed in the LJ units of length,
mass, energy, and time; namely, $\sigma=\sigma_{AA}$, $m=m_{A}$,
$\varepsilon=\varepsilon_{AA}$, and
$\tau=\sigma\sqrt{m/\varepsilon}$. The MD simulations were performed
using the LAMMPS open-source, parallel code~\cite{Lammps} with the
integration time step $\triangle t_{MD}=0.005\,\tau$.

\vskip 0.05in


The first step in the preparation procedure was to arrange all atoms
at the sites of the fcc lattice and then equilibrate the binary
mixture at the temperature $T_{LJ}=1.0\,\varepsilon/k_B$ and zero
pressure. Here, $k_B$ and $T_{LJ}$ indicate the Boltzmann constant
and temperature, respectively.  In all simulations, the system
temperature was regulated via the Nos\'{e}-Hoover
thermostat~\cite{Allen87,Lammps}. Periodic boundary conditions were
imposed along all three dimensions.  Next, the system was linearly
cooled down to an annealing temperature $T_a$ (in the vicinity of
the glass transition temperature) during $10^4\,\tau$ at zero
pressure. Once at $T_a$, the binary mixture was relaxed at constant
temperature and pressure during the time interval
$2\times10^5\,\tau$ (i.e., $4\times10^7$ MD steps). Following the
relaxation period, the system was instantaneously quenched to the
very low temperature $T_{LJ}=0.01\,\varepsilon/k_B$ and then further
relaxed during the additional time interval $10^4\,\tau$ at zero
pressure.

\vskip 0.05in


After the thermal treatment, the mechanical properties were obtained
by imposing shear strain deformation at constant volume and
temperature $T_{LJ}=0.01\,\varepsilon/k_B$. The shear modulus and
the peak value of the stress overshoot were computed from the
stress-strain curves, and the data were averaged over 15 independent
samples. The deformation procedure was repeated for different values
of the strain rate.  This required substantial computational
resources (about 1000 processors).  In all simulations, the system
temperature, potential energy, stress components, and atomic
configurations were periodically saved for the post-processing
analysis.

\section{Results}
\label{sec:Results}


It has been long realized that physical and mechanical properties of
metallic glasses depend crucially on the annealing
history~\cite{Greer16}. An auxiliary concept to characterize the
structural degrees of freedom is the so-called \textit{fictive
temperature} defined as the temperature at which the glass would be
in equilibrium if suddenly brought to it from its current
state~\cite{Tool46,Richter16}. In the limiting case of an infinitely
fast quenching rate, used in the present study, the fictive
temperature essentially coincides with the annealing temperature at
which the system is equilibrated before the quench. Interestingly,
it was demonstrated experimentally for different metallic glasses
that an abrupt toughening transition between brittle and ductile
regimes occurs as functions of the fictive temperature and strain
rate~\cite{Schroers18}.   It was argued that the mechanical
transition originates from the competition between two time scales;
the loading time, which is inversely proportional to the applied
strain rate, and the plastic deformation time scale, determined by
the density of plasticity carriers or shear transformation
zones~\cite{Schroers18}.

\vskip 0.05in


In the recent MD study, it was found that the glass transition
temperature of the KA mixture of $N=60\,000$ particles at zero
pressure is $T_g\approx0.35\,\varepsilon/k_B$~\cite{Priez19one}.
This value was obtained by linearly extrapolating the low and high
temperature dependence of the potential energy during gradual
cooling with the computationally slow rate of
$10^{-5}\varepsilon/k_{B}\tau$.   Note that higher cooling rates
generally result in an increase of the glass transition temperature,
which can be understood from the assumption that the relaxation time
of the system follows a Vogel-Fulcher dependence on
temperature~\cite{Vollmayr96}.   For reference, the mode-coupling
critical temperature of the KA model at the fixed density
$\rho=1.2\,\sigma^{-3}$ is $T_c=0.435\,\varepsilon/k_B$, which was
determined from the fit of the diffusion coefficients of either $A$
or $B$ types of particles to the power-law functions of temperature
at constant volume~\cite{KobAnd95}.

\vskip 0.05in

\begin{figure}[t]
\includegraphics[width=12.0cm,angle=0]{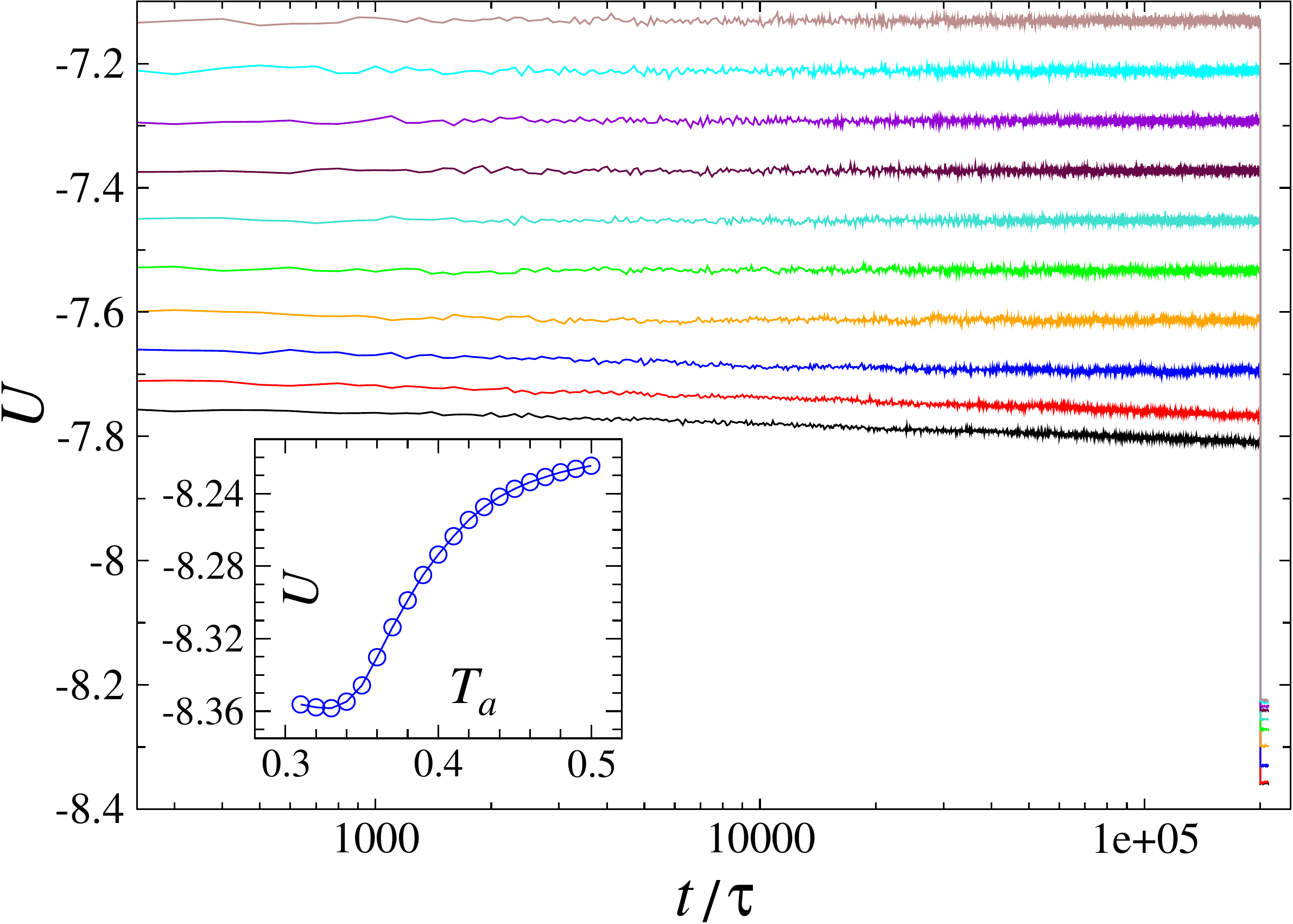}
\caption{(Color online) The time dependence of the potential energy,
$U/\varepsilon$, at various annealing temperatures, $T_a$, and after
the quench to $T_{LJ}=0.01\,\varepsilon/k_B$ at $P=0$. The values of
the annealing temperature are $T_a\,k_B/\varepsilon=0.32$, $0.34$,
$0.36$, $0.38$, $0.40$, $0.42$, $0.44$, $0.46$, $0.48$, and $0.50$
(from bottom to top). The inset shows the potential energy at
$T_{LJ}=0.01\,\varepsilon/k_B$ after the time interval of
$10^4\,\tau$. The data in the inset are averaged over 15 samples.}
\label{fig:poten_age_quench}
\end{figure}

In our study, the system was relaxed at zero pressure in the range
of annealing temperatures, $0.31\,\varepsilon/k_B \leqslant T_a
\leqslant 0.50\,\varepsilon/k_B$, near the glass transition
temperature. The variation of the potential energy per particle
during the relaxation time interval of $2\times10^5\,\tau$ and a
subsequent quench to $T_{LJ}=0.01\,\varepsilon/k_B$ is shown in
Fig.\,\ref{fig:poten_age_quench} for selected values of the
annealing temperature.  It can be observed that the potential energy
remains nearly constant at $T_a \gtrsim 0.38\,\varepsilon/k_B$,
while it decreases gradually during aging at lower temperatures. To
estimate the degree of structural relaxation during
$2\times10^5\,\tau$, we computed the mean square displacement (MSD)
of atoms at low annealing temperatures.  For example, the averaged
MSD is $r^2\approx1.64\,\sigma^2$ at $T_a = 0.32\,\varepsilon/k_B$,
indicating that each atom undergoes a number of cage jumps during
the relaxation time interval.  This estimate is based on two atomic
configurations separated by the time interval $2\times10^5\,\tau$.

\vskip 0.05in


After the system was equilibrated at a certain temperature near
$T_g$, it was instantaneously quenched to the test temperature
$T_{LJ}=0.01\,\varepsilon/k_B$ and then relaxed during $10^4\,\tau$
at zero pressure.  This procedure aims at freezing out the thermal
vibrations but keeping intact the average glass structure. The inset
in Fig.\,\ref{fig:poten_age_quench} shows the dependence of the
potential energy after $10^4\,\tau$ at
$T_{LJ}=0.01\,\varepsilon/k_B$ as a function of the annealing
temperature. It can be seen that the potential energy, which is
mostly determined by the glass structure, decreases towards a
minimum when $T_a$ approaches the glass transition temperature from
above. These results confirm that the potential energy of inherent
structures becomes lower when the temperature is reduced towards the
glass transition temperature~\cite{Stillinger84,Sastry13}. Test
simulations at lower annealing temperatures, $0.28\,\varepsilon/k_B
\leqslant T_a \leqslant 0.30\,\varepsilon/k_B$, have shown that $U$,
measured at $T_{LJ}=0.01\,\varepsilon/k_B$, continues to increase
upon further reducing $T_a$ (not shown). This trend can be explained
by realizing that the effective cooling rate increases when the
system temperature is reduced from $T_{LJ}=1.0\,\varepsilon/k_B$ to
lower values of $T_a$ during the fixed time interval of
$10^4\,\tau$. This way the system is annealed to a glass phase with
higher energy states.

\vskip 0.05in

\begin{figure}[t]
\includegraphics[width=12.0cm,angle=0]{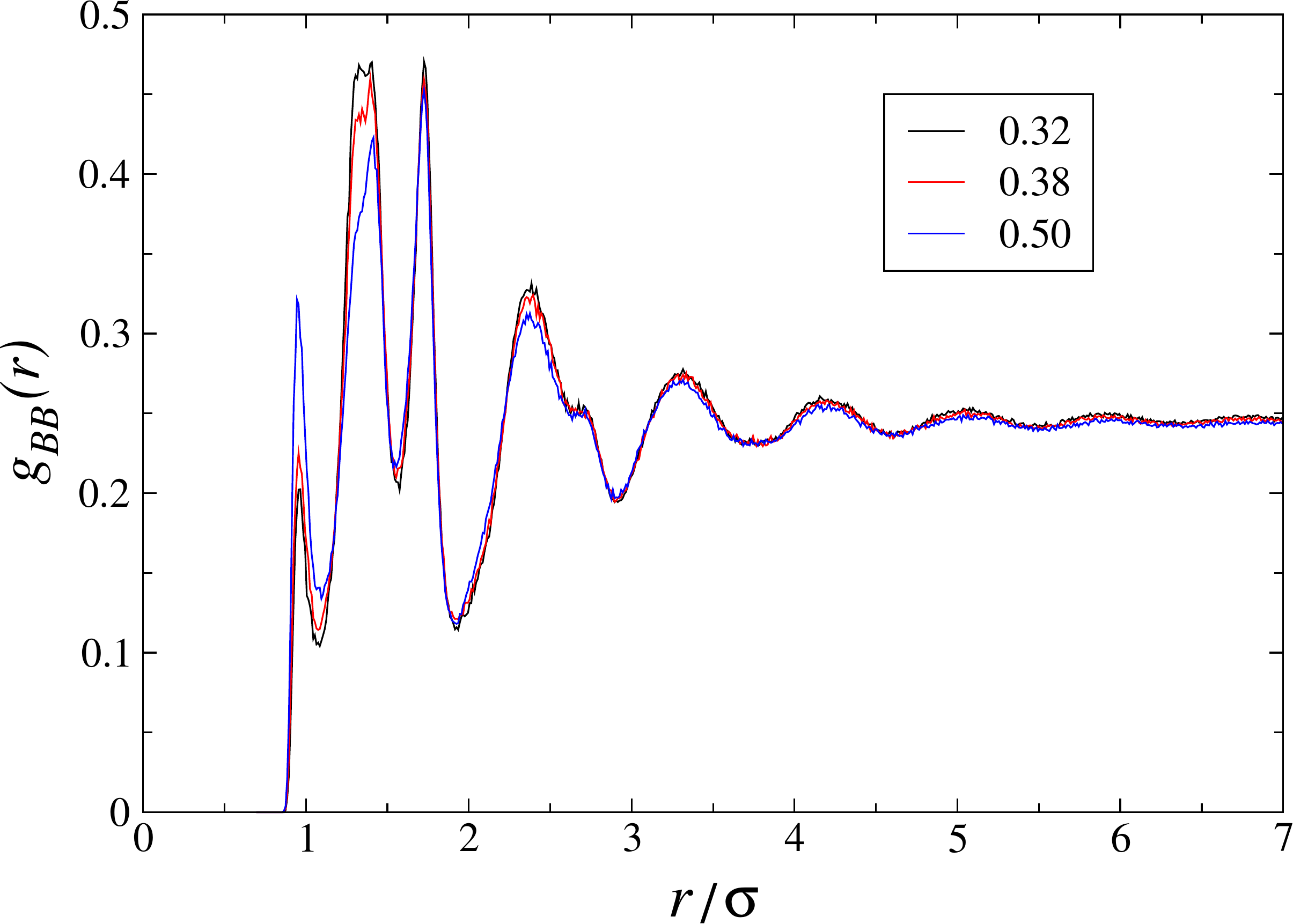}
\caption{(Color online) The pair distribution function of smaller
atoms of type $B$, $g_{BB}(r)$, for glasses prepared at different
annealing temperatures. The values of $T_a$ (in units of
$\varepsilon/k_B$) are listed in the legend. The distribution
function was computed after the samples were instantaneously
quenched from $T_a$ to $T_{LJ}=0.01\,\varepsilon/k_B$ and then
relaxed during the time interval $10^4\,\tau$ at $P=0$. The data are
averaged over 15 independent samples for each value of $T_a$. }
\label{fig:grBB}
\end{figure}

We next analyze the changes in the glass structure for systems
prepared at different annealing temperatures.  It was previously
realized that the most sensitive measure of the atomic structure of
the KA model glass is the shape of the pair correlation function of
the smaller atoms of type $B$~\cite{Vollmayr96,Stillinger00}. This
conclusion stems from the fact that the interaction energy,
$\varepsilon_{BB}$ in Eq.\,(\ref{Eq:LJ_KA}), is the lowest among the
three energies, and, therefore, the number of nearest-neighbor
contacts between smaller, more mobile, atoms is reduced at lower
temperatures~\cite{Stillinger00}.   In our study, the averaged pair
distribution function is plotted in Fig.\,\ref{fig:grBB} for systems
prepared at $T_a = 0.32\,\varepsilon/k_B$, $0.38\,\varepsilon/k_B$,
and $0.50\,\varepsilon/k_B$.  It can be readily observed that the
height of the first peak is significantly reduced when the annealing
temperature is varied from $0.50\,\varepsilon/k_B$ to
$0.32\,\varepsilon/k_B$, while the opposite trend is seen for the
second peak.   We comment that the variation of the first peak
height in Fig.\,\ref{fig:grBB} is much more pronounced than the
change reported during the heat treatment process, where the glass
structure was reset by heating above $T_g$ and then the system was
cooled back to the glass phase with different rates, resulting in
slightly rejuvenated states~\cite{Priez19one}.

\vskip 0.05in

\begin{figure}[t]
\includegraphics[width=12.0cm,angle=0]{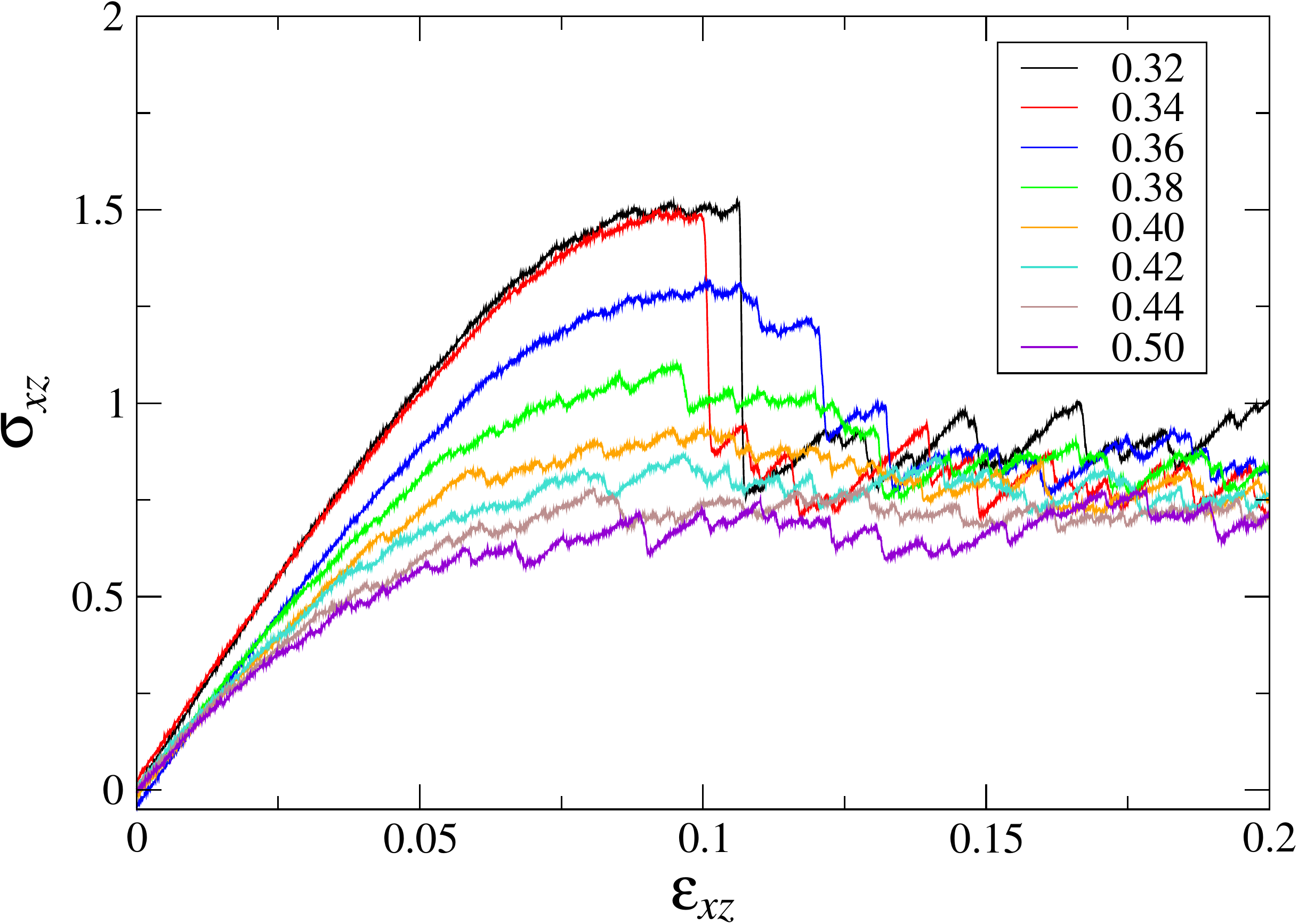}
\caption{(Color online) The variation of shear stress, $\sigma_{xz}$
(in units of $\varepsilon\sigma^{-3}$), as a function of strain,
$\varepsilon_{xz}$, for the indicated values of the annealing
temperature (see the legend). The samples were strained along the
$xz$ plane with the strain rate
$\dot{\varepsilon}_{xz}=10^{-5}\,\tau^{-1}$ at
$T_{LJ}=0.01\,\varepsilon/k_B$ and constant volume. }
\label{fig:stress_strain_rem5}
\end{figure}

The examples of stress-strain curves during shear deformation with
the strain rate $\dot{\varepsilon}_{xz}=10^{-5}\,\tau^{-1}$ are
presented in Fig.\,\ref{fig:stress_strain_rem5} for selected values
of $T_a$.  The deformation was imposed at constant volume and
temperature $T_{LJ}=0.01\,\varepsilon/k_B$ up to $20\%$ shear
strain.  As is evident, the stress response gradually changes from a
monotonic increase and saturation at $T_a = 0.50\,\varepsilon/k_B$
to a pronounced overshoot followed by an abrupt stress drop at $T_a
= 0.32\,\varepsilon/k_B$.   Such behavior typically reflects either
a uniform plastic flow of poorly annealed glasses or the formation
of a shear band in well annealed samples (see discussion below).

\vskip 0.05in

\begin{figure}[t]
\includegraphics[width=12.0cm,angle=0]{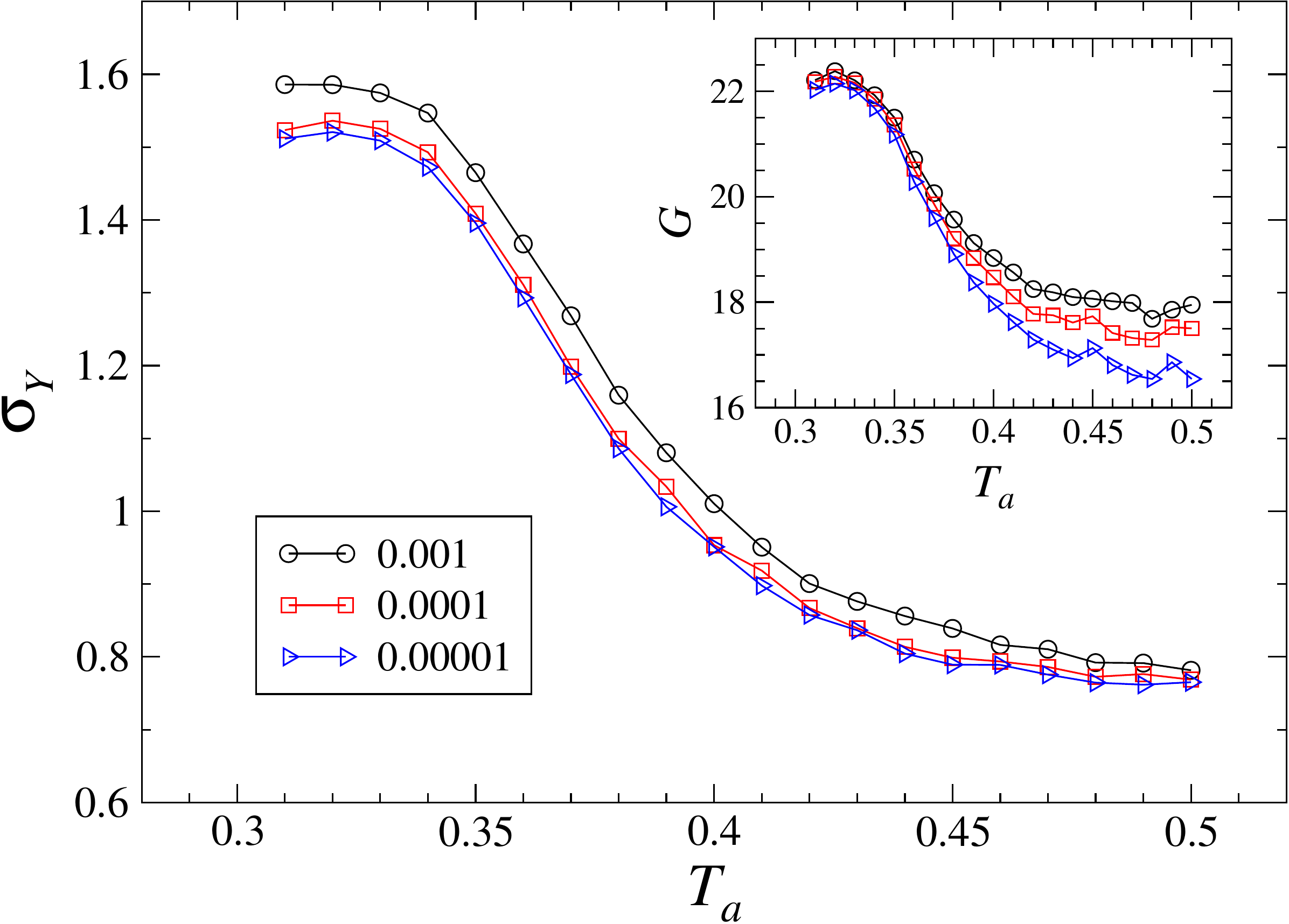}
\caption{(Color online) The peak value of the stress overshoot
$\sigma_Y$ (in units of $\varepsilon\sigma^{-3}$) as a function of
the annealing temperature for the indicated values of the strain
rate, $\dot{\varepsilon}_{xz}$ (in units of $\tau^{-1}$). The inset
shows the shear modulus $G$ (in units of $\varepsilon\sigma^{-3}$)
versus $T_a$ for the same strain rates. The data for $\sigma_Y$ and
$G$ are averaged over 15 samples. }
\label{fig:Y_G_vs_Tf}
\end{figure}

The stress-strain curves were used to estimate the peak value of the
stress overshoot as a function of $T_a$. In turn, the shear modulus
was computed from the slope of the best linear fit to the data at
$\varepsilon_{xz}\leqslant0.01$. Both $\sigma_Y$ and $G$, averaged
over 15 samples, are plotted in Fig.\,\ref{fig:Y_G_vs_Tf} for the
indicated strain rates.   It can be clearly seen that the mechanical
properties depend strongly on the annealing temperature and the
transition occurs when $T_a\approx T_g$.   Furthermore, it was
recently shown experimentally that a toughening transition as a
function of strain rate becomes more pronounced when the fictive
temperature is slightly higher than $T_g$~\cite{Schroers18}. In our
MD setup, the yielding peak becomes higher with increasing strain
rate, as expected; but the increase in $\sigma_Y$ is rather
insensitive to $T_a$ (see Fig.\,\ref{fig:Y_G_vs_Tf}). On the other
hand, as shown in the inset to Fig.\,\ref{fig:Y_G_vs_Tf}, the shear
modulus is a strong function of strain rate only when $T_a > T_g$.
In this sense, the results of MD simulations are consistent with the
conclusions of the experimental study~\cite{Schroers18}.  It should
also be commented that it is difficult to access strain rates
outside the range $10^{-5} \leqslant \dot{\varepsilon}_{xz}\tau
\leqslant 10^{-3}$ reported in Fig.\,\ref{fig:Y_G_vs_Tf}.
Specifically, the averaging becomes computationally expensive at
lower strain rates, while temperature oscillations are detected at
higher strain rates.

\vskip 0.05in


The microscopic details of the deformation process can be captured
via the analysis of the so-called nonaffine displacements of
atoms~\cite{Falk98}.  It is well realized that the local
displacement of an atom with respect to its neighbors can be
computed via the transformation matrix $\mathbf{J}_i$, which
linearly transforms a group of atoms during the time interval
$\Delta t$ and at the same time minimizes the following quantity:
\begin{equation}
D^2(t, \Delta t)=\frac{1}{N_i}\sum_{j=1}^{N_i}\Big\{
\mathbf{r}_{j}(t+\Delta t)-\mathbf{r}_{i}(t+\Delta t)-\mathbf{J}_i
\big[ \mathbf{r}_{j}(t) - \mathbf{r}_{i}(t)    \big] \Big\}^2,
\label{Eq:D2min}
\end{equation}
where the summation is carried over the neighbors within a sphere of
radius $1.5\,\sigma$ with the center at the position vector
$\mathbf{r}_{i}(t)$. It was demonstrated that the nonaffine measure
is an excellent diagnostic of localized shear transformations in
deformed amorphous materials~\cite{Falk98}. Thus, recent studies on
periodic deformation of disordered solids have shown that the
amplitude of the nonaffine measure is approximately power-law
distributed, while the majority of atoms rearrange reversibly during
one or several subyield cycles~\cite{Priezjev16,Priezjev16a}. Above
the yielding point, the temporal evolution of spatial distributions
of atoms with large nonaffine displacements illustrated the
existence of a transient regime of periodic deformation, followed by
the formation of shear bands in both poorly~\cite{Priezjev18a} and
well~\cite{Priezjev17} annealed binary glasses.   More recently, the
analysis of nonaffine displacements was used to study aging and
rejuvenation during elastostatic loading~\cite{PriezELAST19} as well
as the mechanical~\cite{Priezjev18,Priezjev18a,NVP18strload} and
thermal~\cite{Priez19tcyc} annealing processes in amorphous alloys.

\vskip 0.05in

\begin{figure}[t]
\includegraphics[width=12.cm,angle=0]{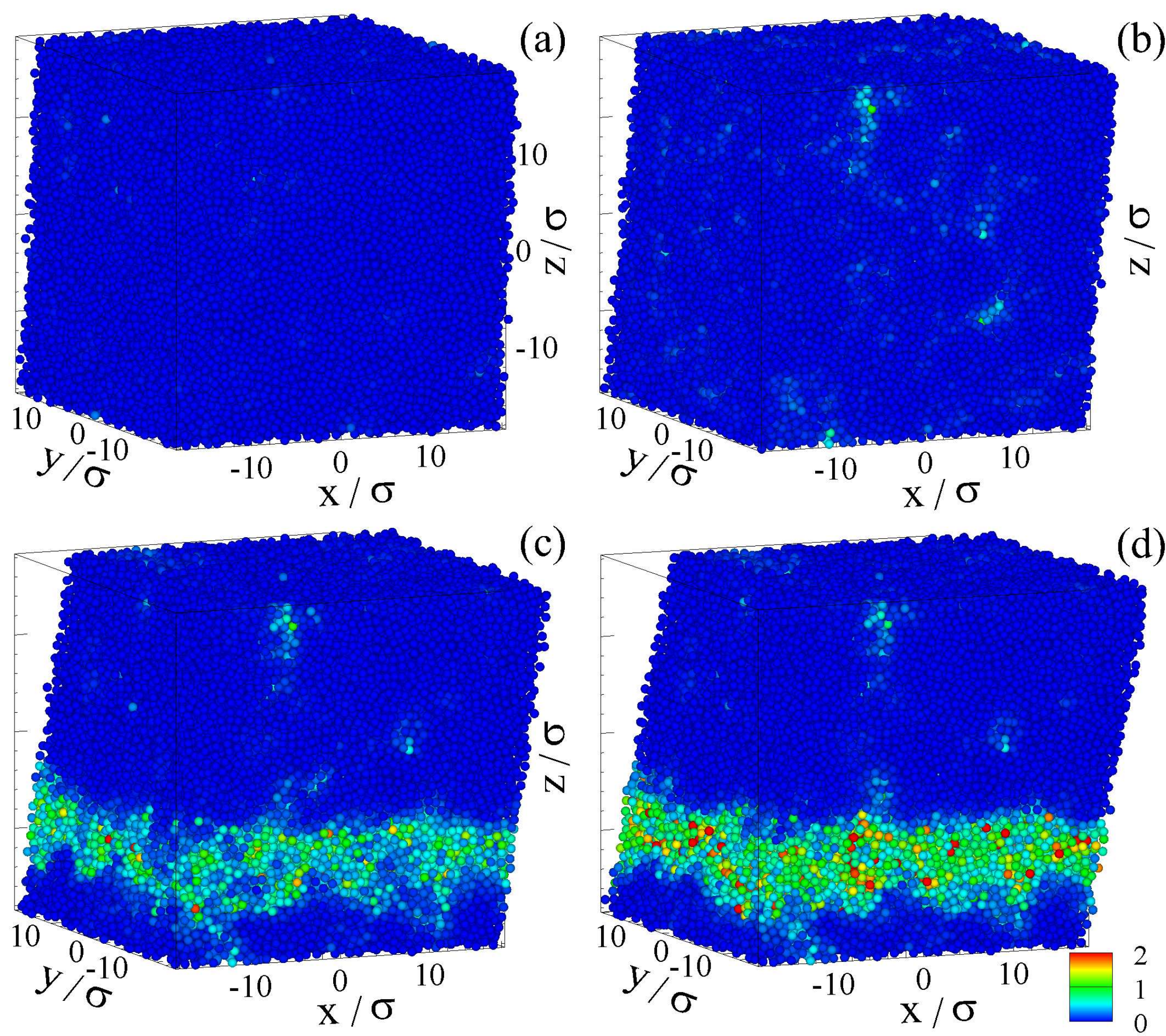}
\caption{(Color online) The sequence of snapshots for the sample
prepared at $T_a=0.32\,\varepsilon/k_B$ and then quenched to
$T_{LJ}=0.01\,\varepsilon/k_B$. The shear strain $\varepsilon_{xz}$
is (a) $0.05$, (b) $0.10$, (c) $0.15$, and (d) $0.20$. The rate of
strain is $\dot{\varepsilon}_{xz}=10^{-5}\,\tau^{-1}$. The color
denotes the nonaffine measure $D^2$ with respect to zero strain (see
the legend). }
\label{fig:snapshots_Tf032}
\end{figure}
\begin{figure}[t]
\includegraphics[width=12.cm,angle=0]{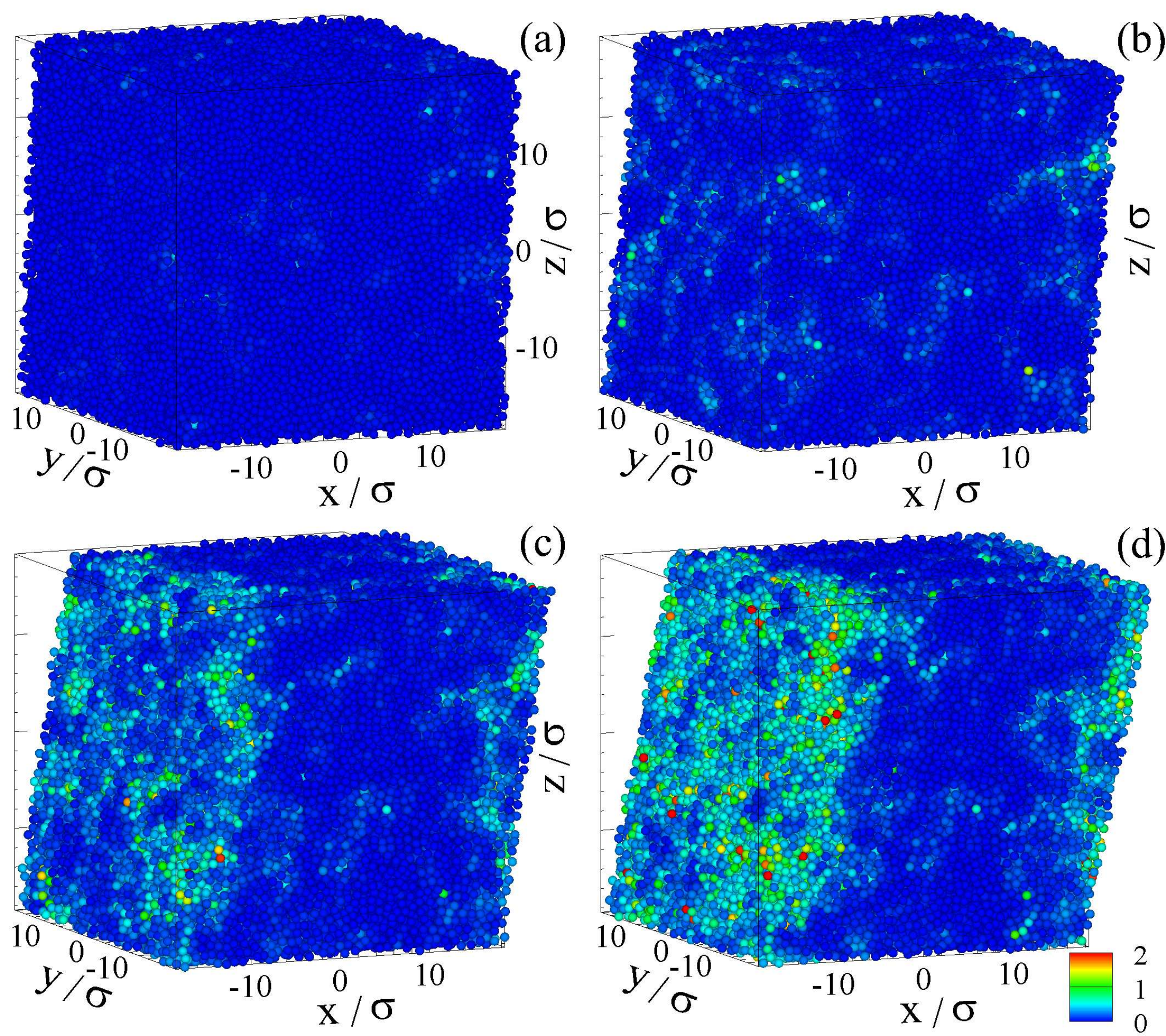}
\caption{(Color online) The series of snapshots depicting atomic
configurations in the strained glass prepared at the annealing
temperature $T_a=0.38\,\varepsilon/k_B$. The shear strain is (a)
$0.05$, (b) $0.10$, (c) $0.15$, and (d) $0.20$. The nonaffine
measure $D^2$ with respect to zero strain is marked according to the
legend. The glass is strained at $T_{LJ}=0.01\,\varepsilon/k_B$ and
constant volume with the rate
$\dot{\varepsilon}_{xz}=10^{-5}\,\tau^{-1}$. The atoms are not drawn
to scale.}
\label{fig:snapshots_Tf038}
\end{figure}
\begin{figure}[t]
\includegraphics[width=12.cm,angle=0]{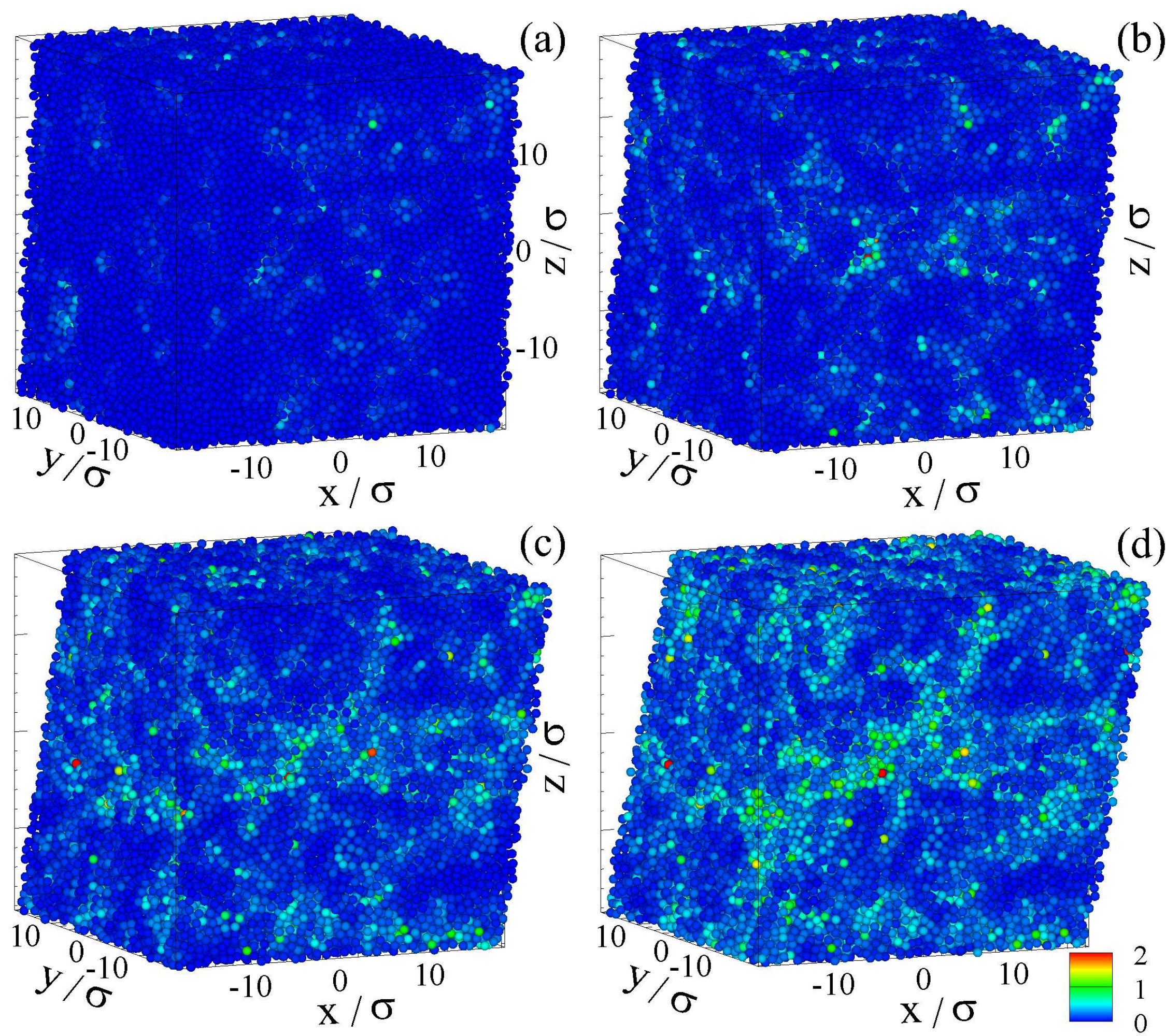}
\caption{(Color online) The consecutive snapshots of the glass
equilibrated at $T_a=0.50\,\varepsilon/k_B$ and then quenched to
$T_{LJ}=0.01\,\varepsilon/k_B$. The strain is (a) $0.05$, (b)
$0.10$, (c) $0.15$, and (d) $0.20$. The nonaffine displacements of
atoms with respect to their neighbors at zero strain are indicated
by the color according to the figure legend. The strain rate is
$\dot{\varepsilon}_{xz}=10^{-5}\,\tau^{-1}$. }
\label{fig:snapshots_Tf050}
\end{figure}

The spatial distribution of nonaffine displacements during steady
shear deformation with the strain rate
$\dot{\varepsilon}_{xz}=10^{-5}\,\tau^{-1}$ is plotted in
Fig.\,\ref{fig:snapshots_Tf032} for $T_a=0.32\,\varepsilon/k_B$, in
Fig.\,\ref{fig:snapshots_Tf038} for $0.38\,\varepsilon/k_B$, and in
Fig.\,\ref{fig:snapshots_Tf050} for $0.50\,\varepsilon/k_B$. The
sequence of instantaneous snapshots are taken at shear strains
$\varepsilon_{xz}=0.05$, $0.10$, $0.15$, and $0.20$, and the
nonaddine measure, Eq.\,(\ref{Eq:D2min}), was computed with respect
to the atomic configuration at zero strain.  It can be clearly seen
in Fig.\,\ref{fig:snapshots_Tf032} that the shear deformation is
accompanied by relatively small clusters of atoms with large
nonaffine displacements at $\varepsilon_{xz}\leqslant0.10$, and the
shear band is formed and become fully developed along the $xy$ plane
at higher strains.  This behavior is consistent with the stress
response shown in Fig.\,\ref{fig:stress_strain_rem5} for
$T_a=0.32\,\varepsilon/k_B$, where the stress drop at
$\varepsilon_{xz}\approx 0.11$ is associated with the formation of
the shear band.

\vskip 0.05in


Next, as illustrated in Fig.\,\ref{fig:snapshots_Tf038}, the
deformation pattern remains similar for the sample prepared at the
higher annealing temperature $T_a=0.38\,\varepsilon/k_B$, except
that the shear band appears to be thinner and it is oriented
perpendicular to the plane of shear. Such orientation is allowed due
to periodic boundary conditions, and it was observed in the previous
MD simulations of binary glasses under steady and periodic shear
deformation~\cite{Horbach16,Priezjev17}.  In sharp contrast, a more
homogeneous plastic deformation is detected in glasses equilibrated
at $T_a=0.50\,\varepsilon/k_B$, as shown in
Fig.\,\ref{fig:snapshots_Tf050}. The inherent structure at this
annealing temperature has the highest potential energy level (see
the inset in Fig.\,\ref{fig:poten_age_quench}) and, therefore, the
largest number of shear transformation zones with low instability
thresholds, resulting in homogeneous deformation. Altogether, these
results demonstrate the importance of controlling the annealing
temperature in order to optimize the mechanical properties of bulk
metallic glasses and improve their plasticity.

\section{Conclusions}

In summary, molecular dynamics simulations were carried out to
investigate the influence of the thermal treatment on structure and
mechanical properties of disordered solids. A model glass was
represented by the three-dimensional binary mixture with highly
non-additive parametrization that prevents crystallization near the
glass transition temperature.   The binary mixture was first
equilibrated at a temperature close to the glass transition point
and then instantaneously quenched and further relaxed at a very low
temperature.   It was shown that the potential energy of the
inherent structures is reduced and the height of the
nearest-neighbor peak in the pair distribution function of smaller
atoms becomes lower in samples prepared at lower annealing
temperatures. The results of mechanical tests indicate that both the
shear modulus and the yielding peak increase significantly when the
annealing temperature is reduced towards the glass transition
temperature. We also found that the strain rate dependence of the
shear modulus becomes apparent only at annealing temperatures
greater than the glass transition temperature.  The analysis of
nonaffine displacements illustrated that the shear deformation
changed from brittle to ductile upon increasing annealing
temperature.

\section*{Acknowledgments}

Financial support from the National Science Foundation (CNS-1531923)
is gratefully acknowledged. The article was prepared within the
framework of the HSE University Basic Research Program and funded in
part by the Russian Academic Excellence Project `5-100'. The
molecular dynamics simulations were performed using the LAMMPS
software developed at Sandia National Laboratories~\cite{Lammps}.
The numerical simulations were performed at Wright State
University's Computing Facility and the Ohio Supercomputer Center.


%

%

%
%

%
%

%
%

%
%

%
%

\bibliographystyle{prsty}

\end{document}